\documentclass[pdflatex,sn-mathphys-num,a4paper]{sn-jnl}

\usepackage{url} 
\newcommand{\dNdX}{\ensuremath{dN/dx}} 
\usepackage{graphicx}
\usepackage{multirow}
\usepackage{amsmath,amssymb,amsfonts}
\usepackage{amsthm}
\usepackage{mathrsfs}
\usepackage[title]{appendix}
\usepackage{xcolor}
\usepackage{textcomp}
\usepackage{manyfoot}
\usepackage{booktabs}
\usepackage{algorithm}
\usepackage{algorithmicx}
\usepackage{algpseudocode}
\usepackage{listings}
\usepackage{caption}
\usepackage{float}
\usepackage{adjustbox} 
\usepackage{placeins} 

\theoremstyle{thmstyleone}

\theoremstyle{thmstyletwo}

\theoremstyle{thmstylethree}

\begin{document}

\title[Article Title]{Integration and characterization of Readout Electronics System for dN/dx Measurement with Drift Chamber Prototype}

\author[1,2,3]{\fnm{Dongcheng} \sur{Cai}}\email{caidc@ihep.ac.cn}
\author[1,2,3]{\fnm{Qicai} \sur{Li}}\email{liqicai@ihep.ac.cn}
\author[1,3]{\fnm{Mingyi} \sur{Dong}}
\author[1,2,3]{\fnm{Weile} \sur{Gong}}
\author[1,3]{\fnm{Mengyang} \sur{Ji}}
\author*[1,2,3]{\fnm{Hongbin} \sur{Liu}}\email{hbliu@ihep.ac.cn} 
\author[1,3]{\fnm{Wenyu} \sur{Pan}}
\author[1,3]{\fnm{Linghui} \sur{Wu}}
\author[2,4]{\fnm{Dewei} \sur{Xu}}
\author[1,2,3]{\fnm{Yimie} \sur{Yuan}}
\author[1,3]{\fnm{Hongyu} \sur{Zhang}}
\author[1,3]{\fnm{Guang} \sur{Zhao}}
\author[1,2,3]{\fnm{Yubin} \sur{Zhao}}

\affil[1]{ \orgname{Institute of High Energy Physics, Chinese Academy of Sciences}, \orgaddress{ \city{Beijing}, \postcode{100049}, \country{China}}}
\affil[2]{ \orgname{Spallation Neutron Source Science Center}, \orgaddress{ \city{Dongguan}, \postcode{523803}, \state{Guangdong}, \country{China}}}
\affil[3]{\orgname{University of Chinese Academy of Sciences}, \orgaddress{ \city{Beijing}, \postcode{100049},  \country{China}}}
\affil[4]{ \orgname{Dongguan University of Technology}, \orgaddress{\city{Dongguan}, \postcode{523808}, \state{Guangdong}, \country{China}}}

\abstract{To explore the feasibility of high-precision particle identification using the cluster counting technique for the drift chamber, a dedicated readout electronics system with low noise, high bandwidth, and high sampling rate is required. This paper presents the design and performance evaluation of a scalable readout prototype developed for this application. The system architecture integrates a custom front-end with a $1.3\ \text{GSps}$ waveform sampling backend, implemented within a modular 120-channel framework. Laboratory characterization of the 40-channel prototype demonstrates a $-3$ dB analog bandwidth of $460\ \text{MHz}$ and an Equivalent Noise Input current of $0.81\ \mu\text{A}_\text{rms}$. These specifications are essential for preserving the fast temporal features of ionization signals. Furthermore, the system achieves an intrinsic timing jitter of $0.87\ \text{ns}$, which satisfies the timing precision requirements for drift distance measurement. Joint experiments with a drift chamber prototype using cosmic rays verified the system's capability to resolve discrete ionization peaks within piled-up waveforms. These results confirm that the readout electronics provide the signal fidelity and temporal resolution necessary for future cluster counting algorithm development.}

\keywords{Data acquisition circuits, Readout electronics, Drift chamber detector}

\maketitle

\section{Introduction}\label{sec1}

The Circular Electron Positron Collider (CEPC) is designed to operate as a Higgs and high-luminosity Z factory, conduct $W^+W^-$ threshold scans, and potentially host future top-quark ($t\bar{t}$) studies. Its primary objective is to perform precision measurements of the Standard Model \cite{An_2019, thecepcstudygroup2025cepctechnicaldesignreport}. Achieving these physics goals, particularly for flavor physics and jet reconstruction, demands efficient particle identification (PID) capabilities. Specifically, the tracking system is required to separate $K/\pi$ across a momentum range up to approximately $20\ \text{GeV}/c$ \cite{chekanov2016conceptual, Zhu2022RequirementAF}. Currently, the CEPC baseline gaseous detector employs a Time Projection Chamber (TPC). As an alternative, the drift chamber solution has been proposed, distinguished by its low material budget \cite{CHIARELLO2019464, article}. In particular, the combination of the cluster counting technique ($dN/dx$) with the drift chamber is regarded as a promising approach for high-precision PID.

Gaseous detectors typically utilize specific energy loss ($dE/dx$) for PID. To mitigate the long Landau tails associated with energy deposition, this method generally employs a truncated mean algorithm. However, the statistical fluctuations of the Landau distribution compromise the PID resolution. This leads to reduced $K/\pi$ separation power, especially in the relativistic rise region (momentum $>10\ \text{GeV}/c$) \cite{Zhu2022RequirementAF, Fang_2023}. In response to this limitation, the cluster counting technique has been investigated. Unlike charge integration, $dN/dx$ involves counting individual primary ionization clusters along the particle track. Since the number of primary clusters follows a Poisson distribution, statistical variance is reduced compared to energy loss measurements. Simulation studies indicate that $dN/dx$ improves PID resolution compared to $dE/dx$ for a fixed track length. Theoretically, this technique meets the separation requirements\cite{Fang_2023, Chiarello_2017}.

The experimental implementation of the $dN/dx$ method demands specific performance characteristics from the readout electronics. Extraction of the $dN/dx$ parameter relies on the identification of individual ionization peaks within detector signals, which are subsequently processed by clustering algorithms. While the intrinsic current signal from the drift chamber typically has a rise time of less than $1~\text{ns}$, the signal after preamplification and shaping exhibits a rising edge of approximately $4~\text{ns}$ \cite{cluster_ml}. To preserve these fast transient features without distortion, preamplification with an analog bandwidth exceeding several hundred megahertz is required \cite{dc_electronics}. In the digitization stage, capturing sufficient waveform detail for reconstruction calls for sampling rates in the gigahertz regime \cite{Chiarello_2017}. Furthermore, the minute amplitude of single primary ionization currents imposes strict limits on the system noise floor. Specifically, signals generated by single primary electrons typically carry a total charge of approximately $16~\text{fC}$, assuming a standard gas gain of $10^5$. Given the fast characteristic time of a few nanoseconds, this corresponds to peak currents of approximately $3\text{--}4~\mu\text{A}$. Distinguishing these weak single-electron signals from the electronic background requires controlling the Equivalent Noise Input current (ENI) to approximately $1~\mu\text{A}_\text{rms}$. Additionally, precise drift distance determination mandates timing accuracy better than $1~\text{ns}$. Finally, the architectural design prioritizes scalability and compactness to facilitate deployment in future large-scale experiments \cite{article}. This study presents the development and integration of a readout system tailored for the CEPC drift chamber. The overall 120-channel architecture is described, followed by the performance characterization of a representative 40-channel module. The capability to capture and resolve ionization features is validated through bench tests and cosmic-ray experiments, confirming the system's suitability for future advanced PID studies.

\section{System Design and Integration}\label{sec2}

\begin{figure}[H]
\centering
\includegraphics[width=1\linewidth]{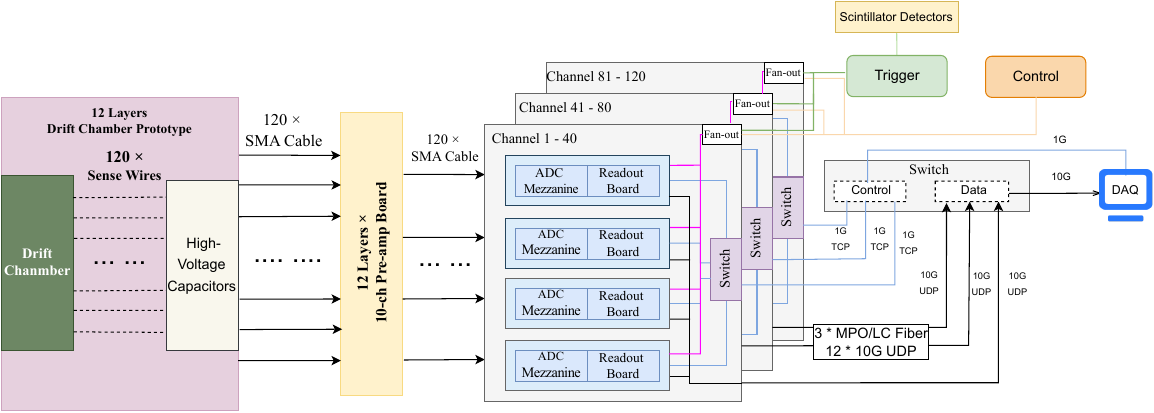} 
\caption{Block diagram of the readout electronics for the drift chamber demonstrator.}\label{fig:readout}
\end{figure}

Fig.~\ref{fig:readout} provides the overview of the component layout and signal transmission path described as follows. 

The analog signals from the drift chamber sense wires are AC-coupled via high-voltage capacitors to block the DC high voltage. The Front-End Electronics (FEE) boards, which consist of the preamplifiers, are installed in a shielded chassis located near the detector. Consequently, the signal transmission distance is reduced, limiting signal attenuation and external interference. Meanwhile, the maintenance access is preserved. The Back-End Electronics (BEE), containing the digitization and readout logic, is placed in a readout crate at a further distance to spatially separate the digital switching circuits from the sensitive analog front-end. Interconnections between the detector, FEE, and BEE are implemented using low-loss SMA cables.

For signal processing, the chain begins with amplification of the induced current signals by the FEE. The conditioned analog signals are then transmitted to the BEE, where they are digitized by AD9695 \cite{ad9695} analog-to-digital converters (ADCs). These ADCs operate at a sampling rate of $1.3\ \text{GSps}$ with a 14-bit vertical resolution, ensuring sufficient temporal resolution for waveform reconstruction. Within the BEE, a Field-Programmable Gate Array (FPGA)-based acquisition module performs data decoding, framing, and buffering. Data transmission to the Data Acquisition (DAQ) server is implemented via two distinct network protocols running over optical and copper links. A $1\ \text{Gbps}$ TCP/IP link over Ethernet is used for system configuration. Meanwhile, a dedicated high-bandwidth data uplink is realized using a $10\ \text{Gbps}$ UDP stream, transmitted from the FPGA's SFP+ transceivers via MPO/LC multi-mode fibers. Both links are aggregated by a network switch and routed to the DAQ server for storage and offline analysis. System-wide synchronization is maintained through externally distributed clock and trigger signals, ensuring time alignment across all 120 channels.

\subsection{Drift Chamber Prototype}\label{subsec2.1}

\begin{figure}[H]
    \centering
    \includegraphics[width=0.9\linewidth]{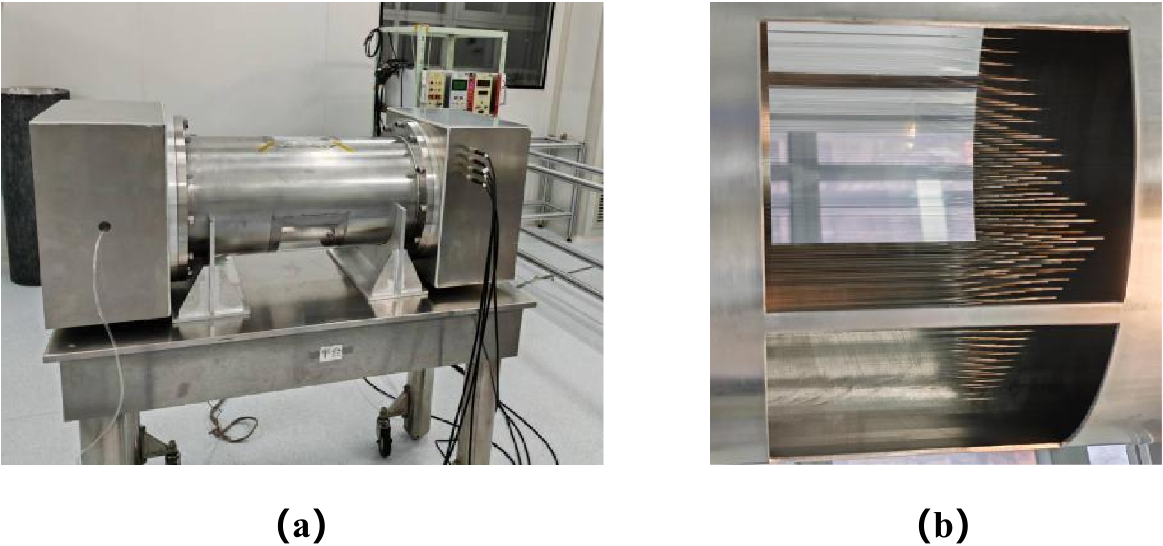}
    \caption{(a) Photograph of the drift chamber prototype. (b) Wire layout of the drift chamber, displaying the precise arrangement of 120 anode sense wires and 416 cathode field wires.}\label{fig_dc_wire}
\end{figure}

The drift chamber prototype consists of multiple layers of precisely arranged thin wires within a gas volume. In the CEPC design, the drift chamber features a cell size of approximately $18\ \text{mm} \times 18\ \text{mm}$ \cite{cluster_ml}. Each cell is centered on an anode sense wire biased at a high positive voltage and surrounded by field-shaping cathode wires illustrated in Fig.~\ref{fig_dc_wire}. As a charged particle traverses the chamber filled with a light gas mixture ($90\%$ $\text{He} + 10\%$ $i\text{C}_4\text{H}_{10}$), it ionizes the gas molecules, creating discrete primary electron-ion pairs along its trajectory.  Driven by the radial electric field, these primary electrons drift toward the central anode. Close to the anode wire, the electrons are multiplied in an avalanche process, and fast transient current signals are induced on the sense wire. These signals typically exhibit rise times on the order of nanoseconds and vary significantly in amplitude due to statistical fluctuations in the ionization process. For the $dN/dx$ method, the capability to resolve these discrete signal peaks is essential. Consequently, the readout electronics need to possess sufficient bandwidth and low noise characteristics to distinguish individual ionization clusters from the background and effectively resolve signal pile-up events caused by the close arrival times of drift electrons.

\subsection{Front-end and Back-end Electronics for dN Measurement}\label{subsec2.2}

In the implementation of the $dN/dx$ measurement method, strict requirements are placed on the readout electronics to extract two physical parameters:  the number of primary ionization clusters ($dN$) and their drift distances ($dx$). $dN$ determination relies on identifying individual, fast-rising current pulses within the composite waveform. Consequently, high-bandwidth analog conditioning and high-resolution digitization are required. Meanwhile, the measurement of $dx$ necessitates precise timing information, which will be discussed in the subsequent trigger section.

\begin{figure}[H]
    \centering
    \includegraphics[width=0.79\linewidth]{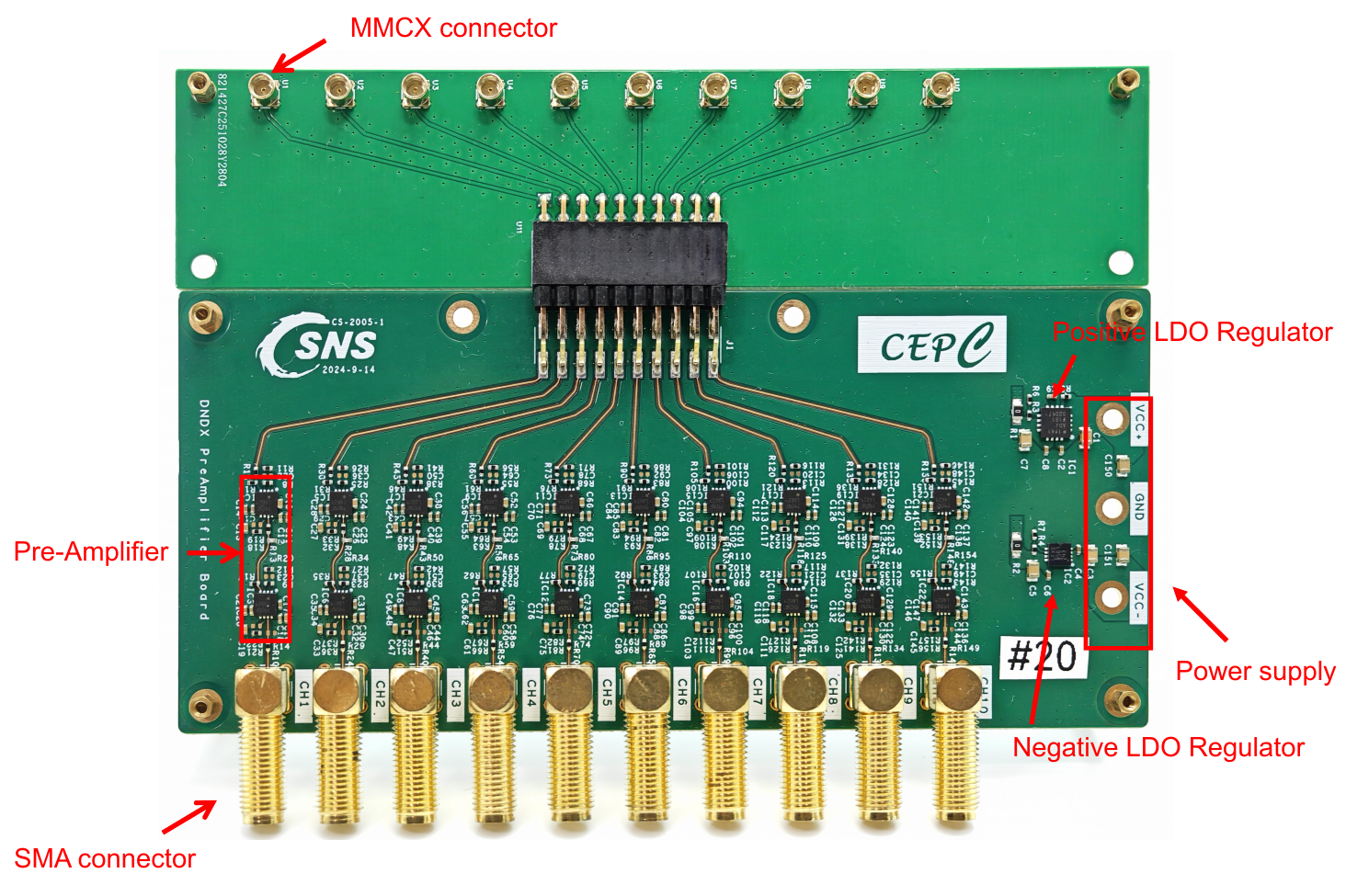}
    \caption{10-channel preamplifier board implemented with $50\ \Omega$ input termination and two-stage voltage amplification.}\label{fig:pre-amp-photo}
\end{figure}

The FEE is implemented as a set of 10-channel preamplifier boards (shown in Fig.~\ref{fig:pre-amp-photo}) mounted close to the chamber to minimize capacitance and noise pickup. For the analog chain, a $50\ \Omega$ input termination is adopted to realize the current-to-voltage conversion. Subsequently, a two-stage high-speed voltage amplifier circuit based on the LMH6629 operational amplifier is developed \cite{lmh6629}. In order to preserve the nanosecond-scale rise time of ionization signals without distortion, the circuit is configured for a bandwidth exceeding $500\ \text{MHz}$. In the design of the preamplifier module, the transimpedance gain is set to approximately $1.25\ \text{k}\Omega$ ($1250\ \text{V/A}$) with a corresponding voltage gain of $25\ \text{V/V}$. These parameters are optimized to amplify the weak ionization currents to match the dynamic range of the backend digitizers while maintaining a low baseline noise floor.

\begin{figure}[H]
    \centering
    \includegraphics[width=1\linewidth]{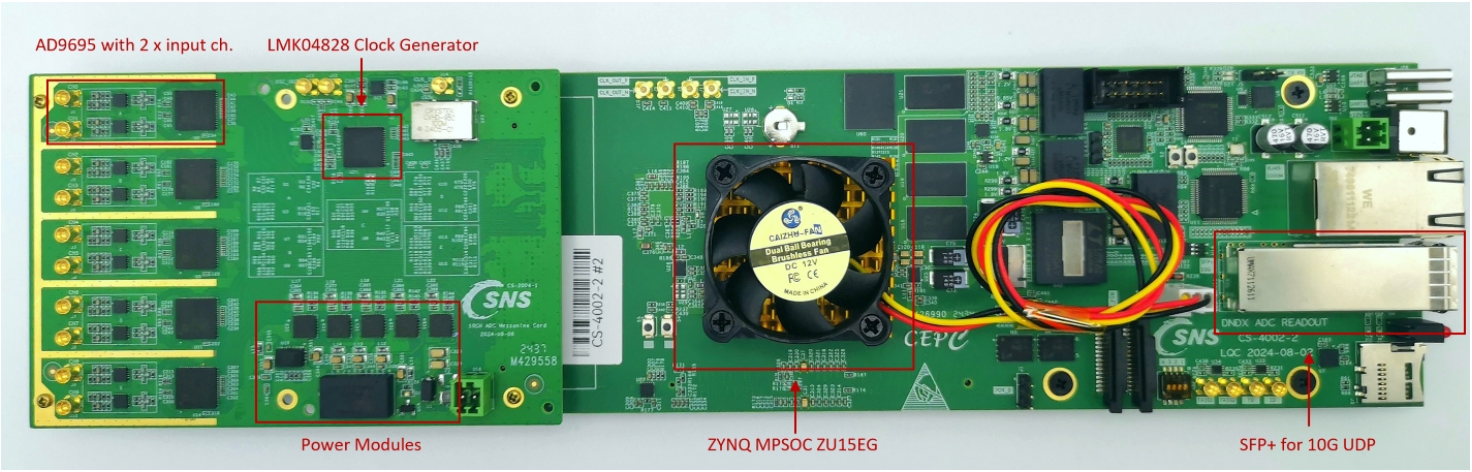}
    \caption{Photograph of the 10-channel readout unit comprising the ADC mezzanine and FPGA carrier.}\label{fig:back-end-photo}
\end{figure}

The amplified signals are processed by the BEE, which adopts a modular architecture. The full 120-channel system is composed of twelve identical 10-channel readout units (shown in Fig.~\ref{fig:back-end-photo}). Each unit consists of an ADC mezzanine card coupled to an FPGA carrier board. The ADC mezzanine integrates high-speed AD9695 converters operating at $1.3\ \text{GSps}$ to capture waveform details. Before digitization, the analog signals undergo single-ended to differential conversion. Taking into account the conversion gain and the loading effect of the ADC input network, the total effective transimpedance gain of the full readout chain is approximately $2232~\Omega$. For the data transfer to the digital logic, the JESD204B protocol is adopted \cite{lmk04828}. The digital core is hosted on the carrier board, featuring an AMD Zynq UltraScale+ ZU15EG FPGA \cite{zynq-ultrascale}. The FPGA utilizes high-speed GTH transceivers (line rates up to $16\ \text{Gbps}$) connected through an FMC interface to receive data from the ADCs. Within the FPGA, firmware modules handle data buffering, trigger alignment, and packetization. The processed waveform data from each unit is transmitted via a dedicated $10\ \text{Gbps}$ UDP link, ensuring sufficient throughput for the high-rate waveform sampling required for cluster counting.

\subsection{Trigger System and Synchronization}\label{subsec2.3}
\begin{figure}[H]
    \centering
    \includegraphics[width=1\linewidth]{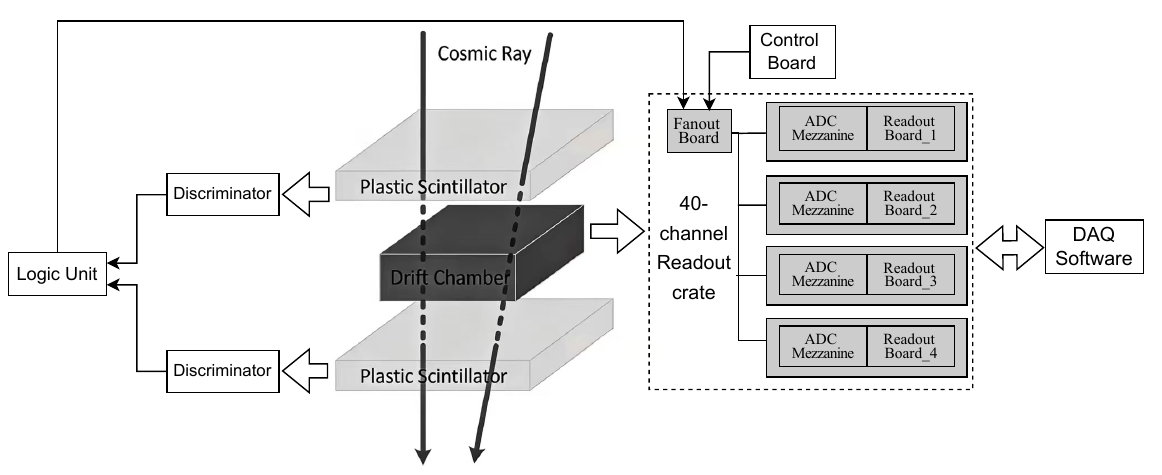}
    \caption{Block diagram of the signal distribution and trigger logic.}\label{platform}
\end{figure}

The trigger system establishes the temporal reference ($t_0$) for the data acquisition and ensures synchronization across the distributed readout modules. The block diagram of the test platform is illustrated in Fig.~\ref{platform}. The primary trigger signal is generated by the coincidence of two 15 cm$\times$15 cm plastic scintillators, which are positioned above and below the drift chamber. The signals from the photomultiplier tubes are processed by discriminators and a logic unit to generate a global trigger when a cosmic ray traverses the fiducial volume. This global trigger is transmitted to the 40-channel readout crate, where it is received by a dedicated fanout board. As shown in the schematic, the fanout board is used as the central synchronization hub. It accepts the trigger signal from the logic unit as well as system command signals from an external Control Board. These timing and control signals are then distributed in parallel to the four readout units installed in the crate. Specifically, the signals are routed to the dedicated trigger input pins on the FPGA carrier board of each unit, establishing the physical link for synchronization.

When the distributed trigger is received, the trigger signal is sampled by the FPGA logic on each readout board based on the local $325\ \text{MHz}$ system clock. Consequently, the ionization signals acquired by all four boards are aligned to a common time base. After digitization, the data is transmitted to the DAQ software for storage and analysis via the bidirectional communication link. For the $dx$ measurement accuracy, the precision of the trigger timestamping is crucial. The detailed characterization of the timing performance is presented in Section~3.4.

\section{Performance of Readout System and Cosmic-Ray Test }\label{sec3}
To evaluate the joint performance of the developed readout electronics and the drift chamber prototype, a laboratory integration platform was established. As the initial phase of system validation, cosmic-ray experiments were conducted to verify the hardware functionality and signal fidelity prior to future beam tests. Since cosmic-ray muons exhibit characteristics of Minimum Ionizing Particles (MIPs), they were adopted as a standard reference source with stable energy deposition. Based on this experimental setup, critical performance metrics were assessed, including the system's linearity, noise performance, and the capability to resolve ionization peaks. 

\begin{figure}[H]
    \centering
    \includegraphics[width=1\linewidth]{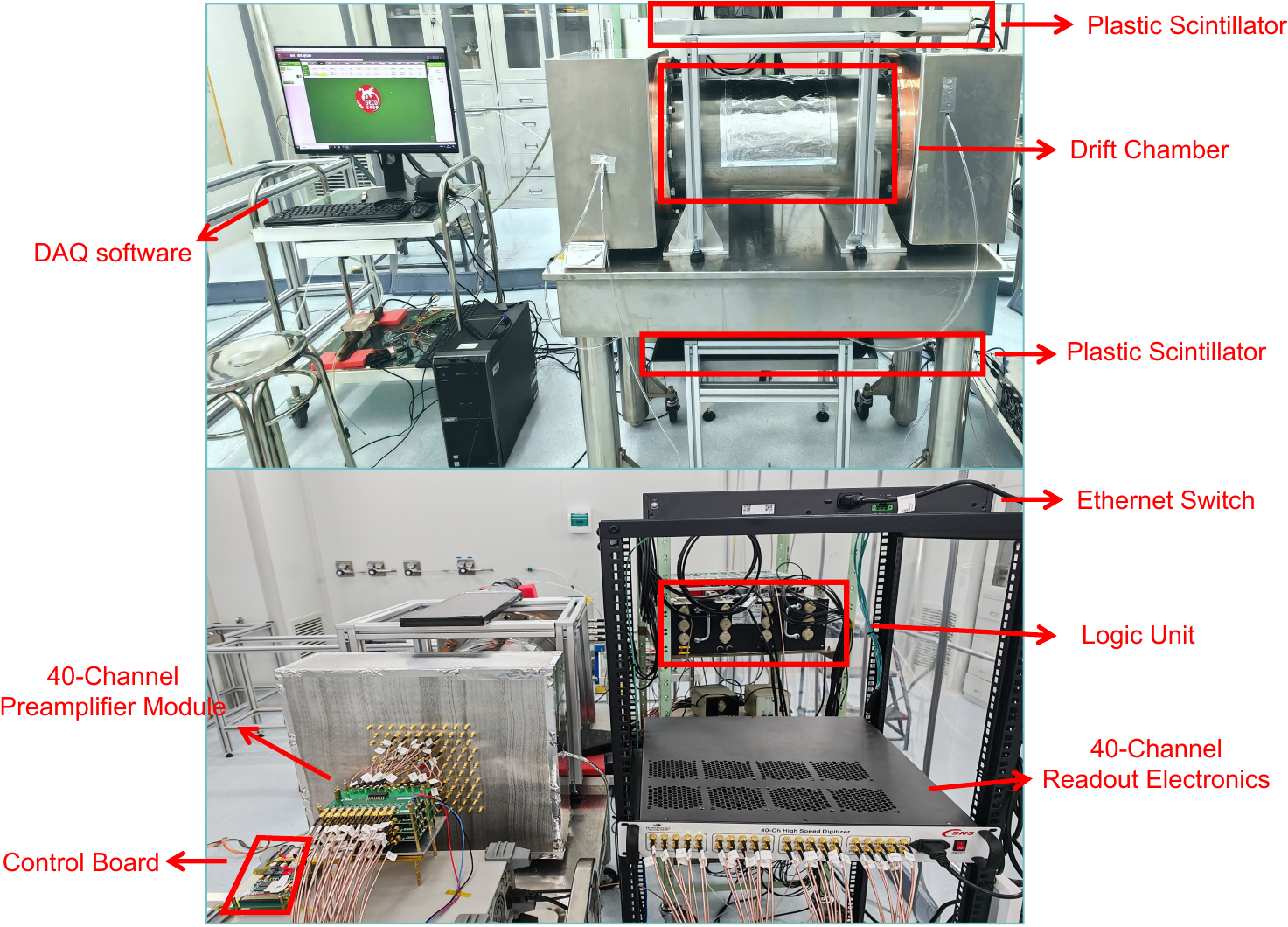}
    \caption{Photograph of the cosmic-ray  experimental setup.}\label{CR_test}
\end{figure}

The photograph of the experimental setup is shown in Fig.~\ref{CR_test}. The test bench is mainly composed of the drift chamber prototype described in Section 2.1. During the tests, the anode sense wires were biased at $1840\ \text{V}$ to ensure operation in the proportional region, yielding a gas gain of approximately $10^5$. For this initial integration phase, the system was instrumented with 40 selected channels to demonstrate the capability of simultaneous waveform recording across multiple layers. Specifically, the connections focused on the central 5 sense wires of 8 selected layers to maximize the hit probability. Signal extraction is achieved via HV decoupling capacitors, with the analog signals subsequently transmitted to the front-end preamplifier module through shielded coaxial cables. Synchronization is managed by a dedicated Trigger and Control Unit, which receives coincidence signals from the scintillator system (as detailed in Section 2.3). Upon validation of a trigger event, the back-end electronics module executes synchronous waveform acquisition and digitization. Finally, the data packets are transmitted via a $10\ \text{Gbps}$ optical uplink to the DAQ server for offline storage and analysis.

\subsection{Linearity and Gain Uniformity}\label{subsec3.1}
For the assessment of the readout electronics, linearity and gain uniformity are essential metrics. The precision of waveform amplitude measurement is directly influenced by these parameters. In order to perform effective peak identification and threshold discrimination, reliable amplitude reconstruction is required.

\begin{figure}[H]
    \centering
    \includegraphics[width=0.7\linewidth]{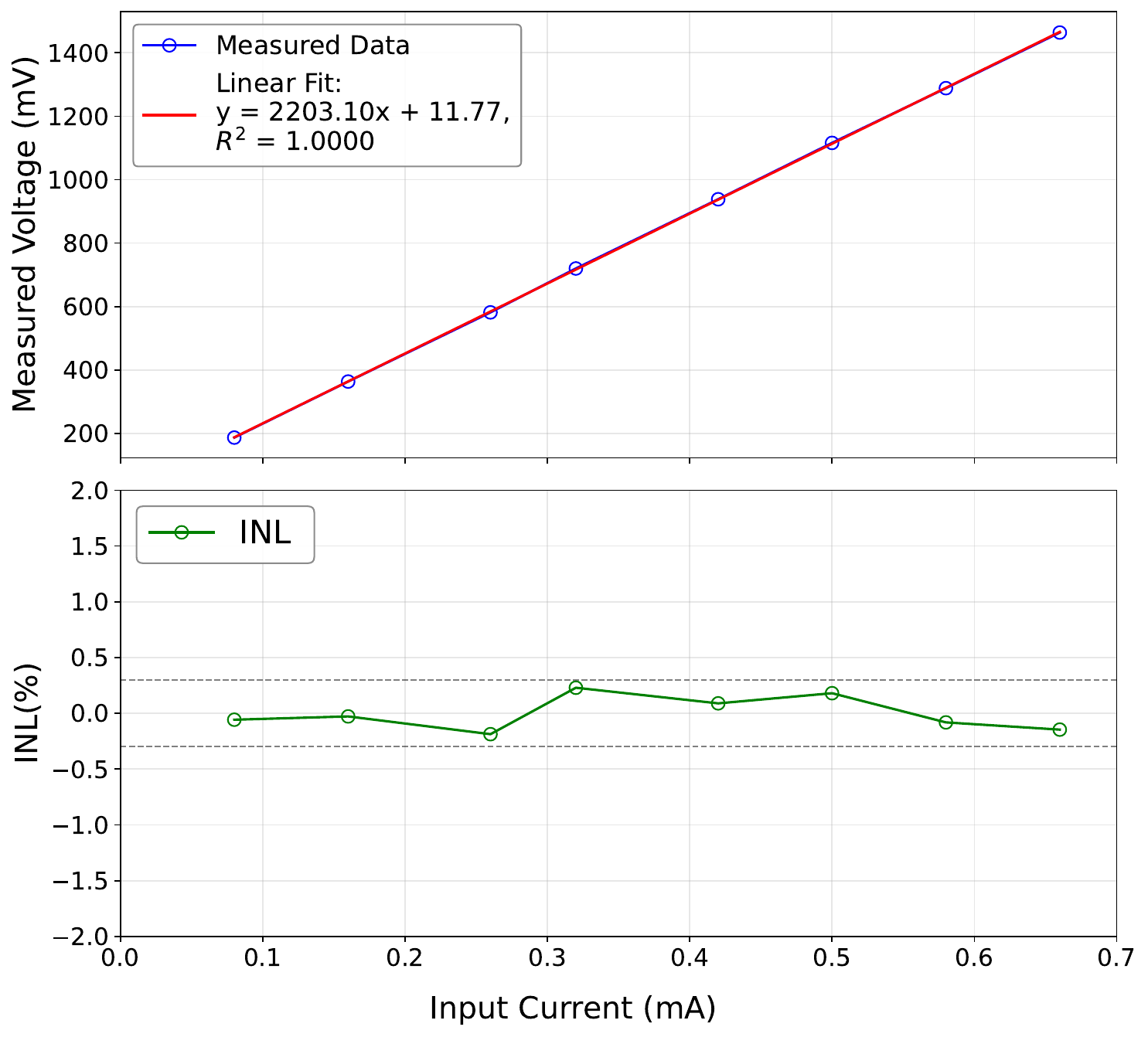}
    \caption{Measured input-output response curve of the readout electronics.}\label{fig:gain_curve}
\end{figure}

To evaluate the amplification linearity, the system response was characterized using sinusoidal inputs corresponding to a current range of $0.08~\text{mA}$ to $0.66~\text{mA}$. This measurement interval was selected to maintain operation within the linear region and avoid signal saturation. The input-output characteristic of the readout chain is shown in Fig.~\ref{fig:gain_curve}. As shown in the top plot, the measured output voltage demonstrates a linear dependence on the input current. Supported by a correlation coefficient $R^2$ of $1.0$, the slope of the linear regression fit indicates an effective transimpedance gain of $2203~\Omega$, which is well consistent with the design value of $2232~\Omega$. The bottom plot displays the normalized residuals, which are distributed within a range of approximately $\pm0.3\%$. The results indicate that the system has stable gain characteristics, which satisfies the requirements for precise \dNdX\ measurement.

\begin{figure}[H]
    \centering
    \includegraphics[width=0.62\linewidth]{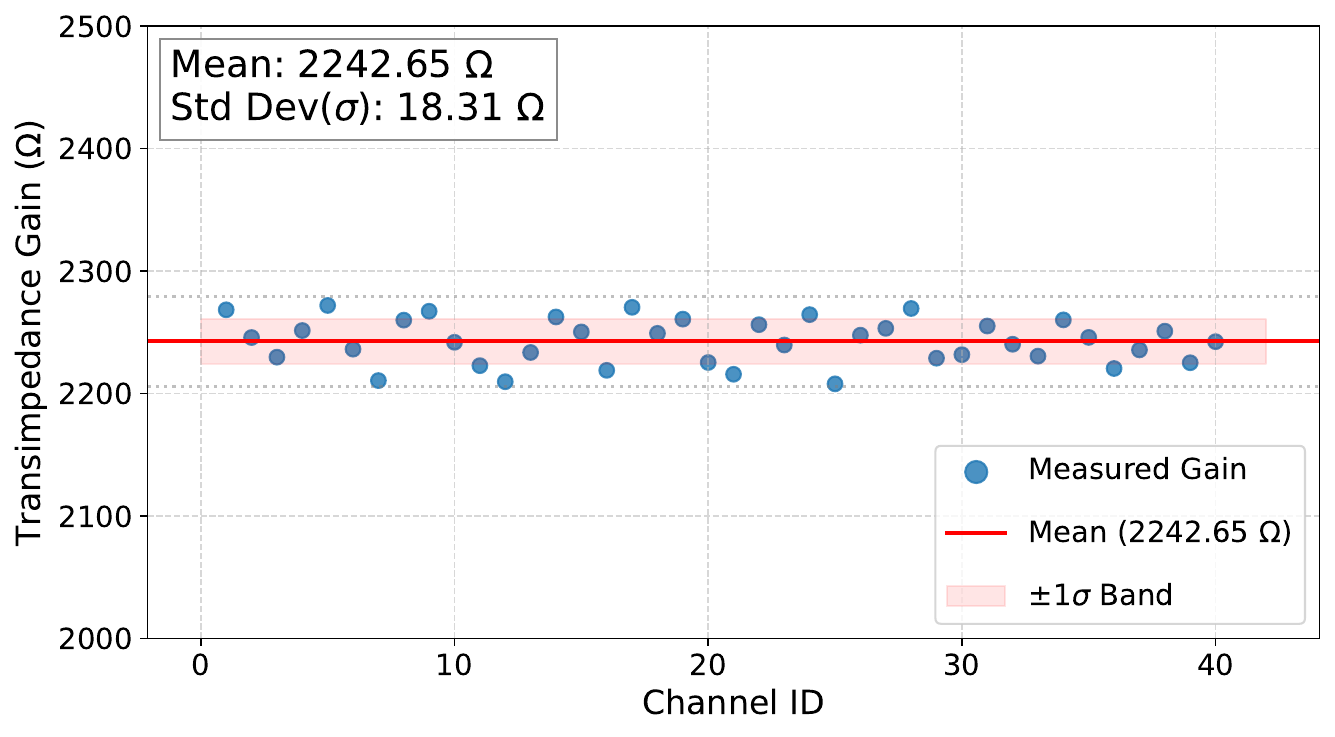}
    \caption{Gain distribution of the 40-channel readout module.}\label{fig:gain_dist}
\end{figure}

In addition, the gain uniformity across the 40-channel module was evaluated to characterize the response uniformity of the system. As presented in the gain distribution (Fig.~\ref{fig:gain_dist}), the 40 channels yield a mean gain of $2242.65\ \Omega$ with a standard deviation ($\sigma$) of $18.31\ \Omega$. The corresponding gain dispersion is calculated to be $0.82\%$. This uniformity facilitates the system calibration process and supports consistent signal response for multi-channel cluster counting tasks.

\subsection{Bandwidth and Pulse Response}\label{subsec3.2}

\begin{figure}[H]
  \centering
  \includegraphics[width=1\linewidth]{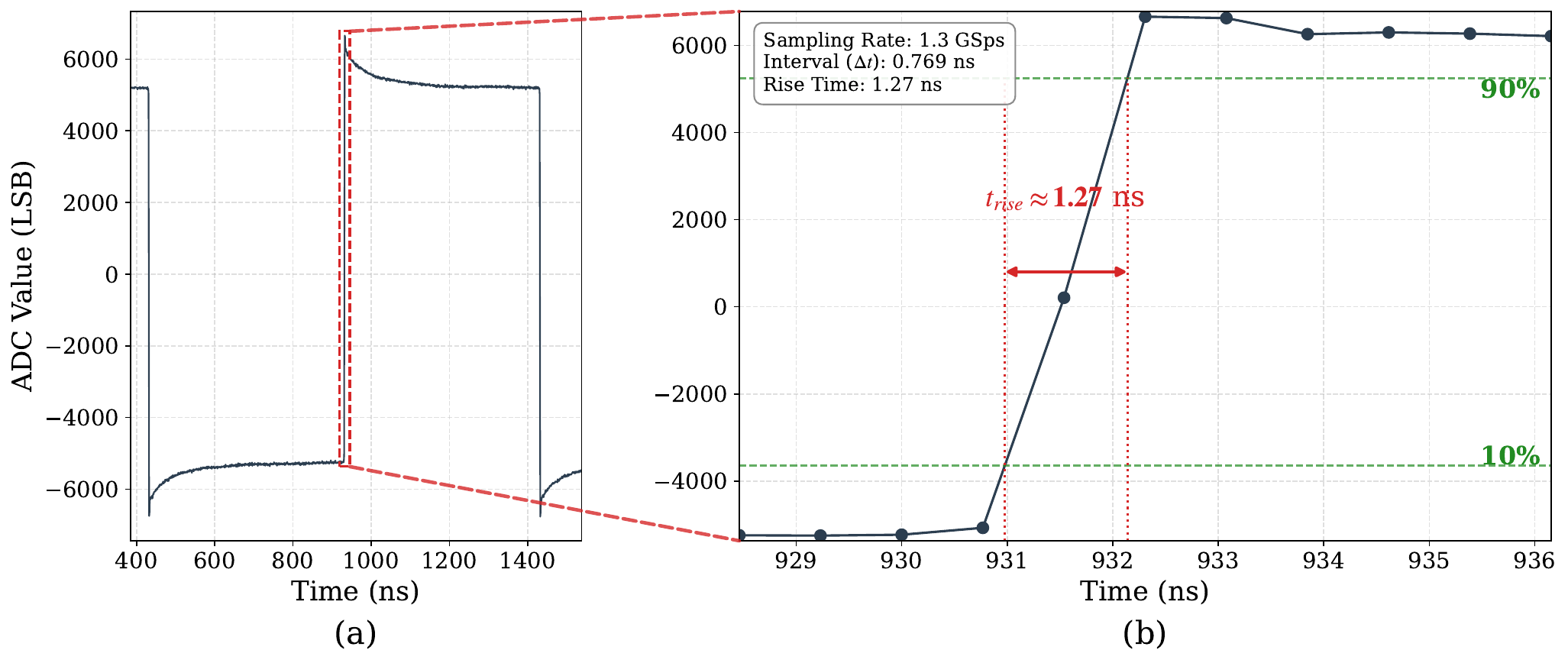} 
  \caption{Digitized waveform response of the readout system to a step pulse input. (a) Full pulse profile; (b) Zoomed-in view of the rising edge.}\label{fig:pulse_response}
\end{figure}

In order to characterize the transient response of the system, a step pulse with a rise time of $t_{\text{r,in}} \approx 1~\text{ns}$ was injected. The digitized waveform is shown in Fig.~\ref{fig:pulse_response}. Plot (a) displays the full pulse profile, while plot (b) provides a zoomed-in view of the rising edge. The system operates at a sampling rate of 1.3 GSps, which imposes a sampling interval of 0.769 ns between consecutive data points. This temporal resolution allows the waveform details to be reconstructed, yielding a measured 10\%–90\% rise time of 1.27 ns. A slight overshoot is observed following the rising edge. However, the amplitude is small and settles rapidly, which indicates that the impact on the signal reconstruction is negligible.

The fast transient response is consistent with the frequency domain characteristics of the system. The $-3\ \text{dB}$ bandwidth of the full readout chain was measured to be approximately $460\ \text{MHz}$. This bandwidth is sufficient for the signal preservation in this application. Based on the formula $t_{\mathrm{r,sys}} \approx 0.35/\mathrm{BW}$, the intrinsic system rise time is estimated to be $0.76\ \text{ns}$. Theoretically, the convolved output rise time is calculated as:
\[
\sqrt{t_{\mathrm{r,in}}^2 + t_{\mathrm{r,sys}}^2} \approx \sqrt{1^2 + 0.76^2} \approx 1.25 \ \mathrm{ns}.
\]
The measured value matches this prediction, confirming the consistency of the bandwidth and timing measurements. Given that the shaped signals from the detector typically exhibit rise times of approximately $4~\text{ns}$ \cite{cluster_ml}, this performance indicates that the readout system is adequate for preserving the leading edges of primary ionization signals required for timing and cluster separation.

\subsection{Pedestal and Noise Characterization}\label{subsec3.3}

A low noise floor facilitates the detection of weak primary ionization clusters. To evaluate the intrinsic noise performance and the impact of detector integration, pedestal measurements were conducted under two conditions: with unconnected inputs and with the drift chamber connected.

\begin{figure}[H]
    \centering
    \includegraphics[width=0.8\linewidth]{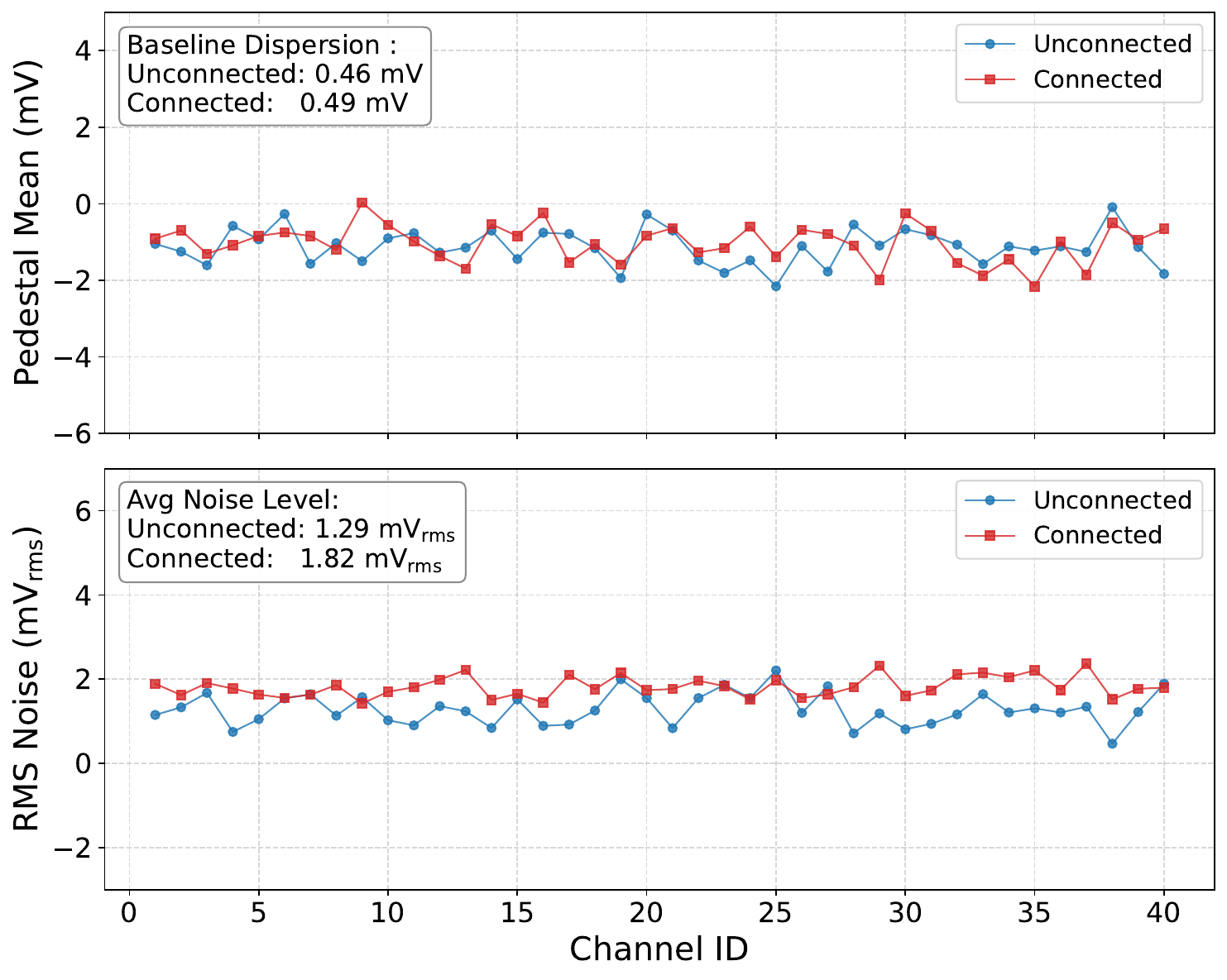}
    \caption{Baseline and RMS noise statistics of the 40-channel system.}\label{fig:baseline}
\end{figure}

The comparison of the baseline characteristics is shown in Fig.~\ref{fig:baseline}. The top plot illustrates the pedestal mean distribution for both scenarios. The data shows baseline uniformity with a channel-to-channel dispersion of 0.46 mV for the unconnected state and 0.49 mV after connecting the detector. This comparison indicates that the baseline remains stable upon detector integration. The bottom plot depicts the RMS noise level for each channel. The average noise level is measured to be $1.29~\mathrm{mV}_\text{rms}$ for the unconnected configuration and $1.82~\mathrm{mV}_\text{rms}$ with the detector connected. While a slight increase in noise is observed due to the added detector capacitance, the noise performance remains stable. Based on the measured transimpedance gain of $2242~\Omega$, the noise level with the detector connected corresponds to the ENI of approximately $0.81~\mu\text{A}_\text{rms}$. This performance meets the design requirement of $1~\mu\text{A}_\text{rms}$, confirming that the system possesses the sensitivity necessary for resolving single primary ionization electrons.

\subsection{Timing Precision and Spatial Resolution Contribution}
\begin{figure}[H]
    \centering
    \includegraphics[width=0.66\linewidth]{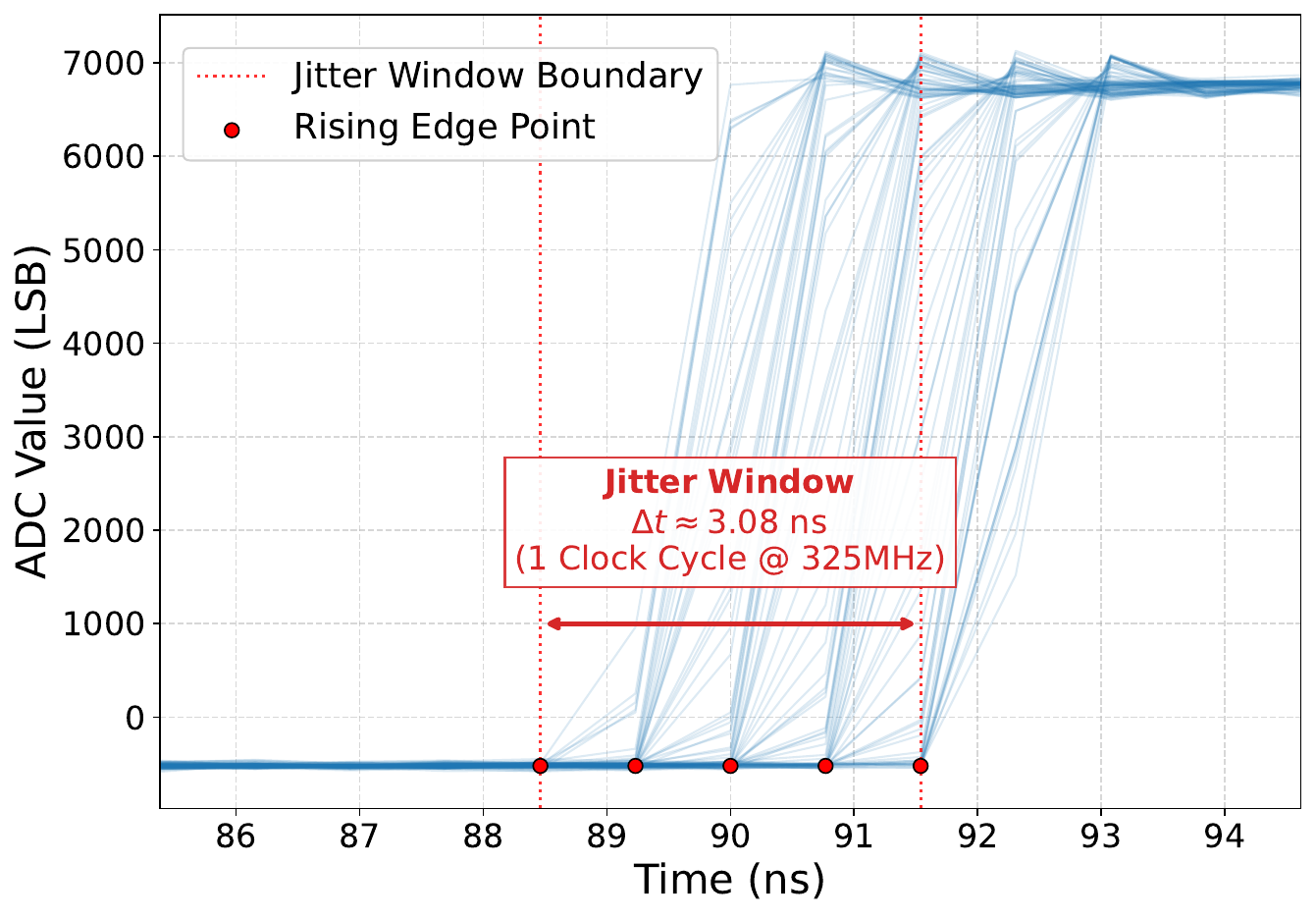}
    \caption{Waveform superposition of 100 asynchronous trigger events.}\label{fig:jitter_superposition}
\end{figure}

\begin{figure}[H]
    \centering
    \includegraphics[width=0.6\linewidth]{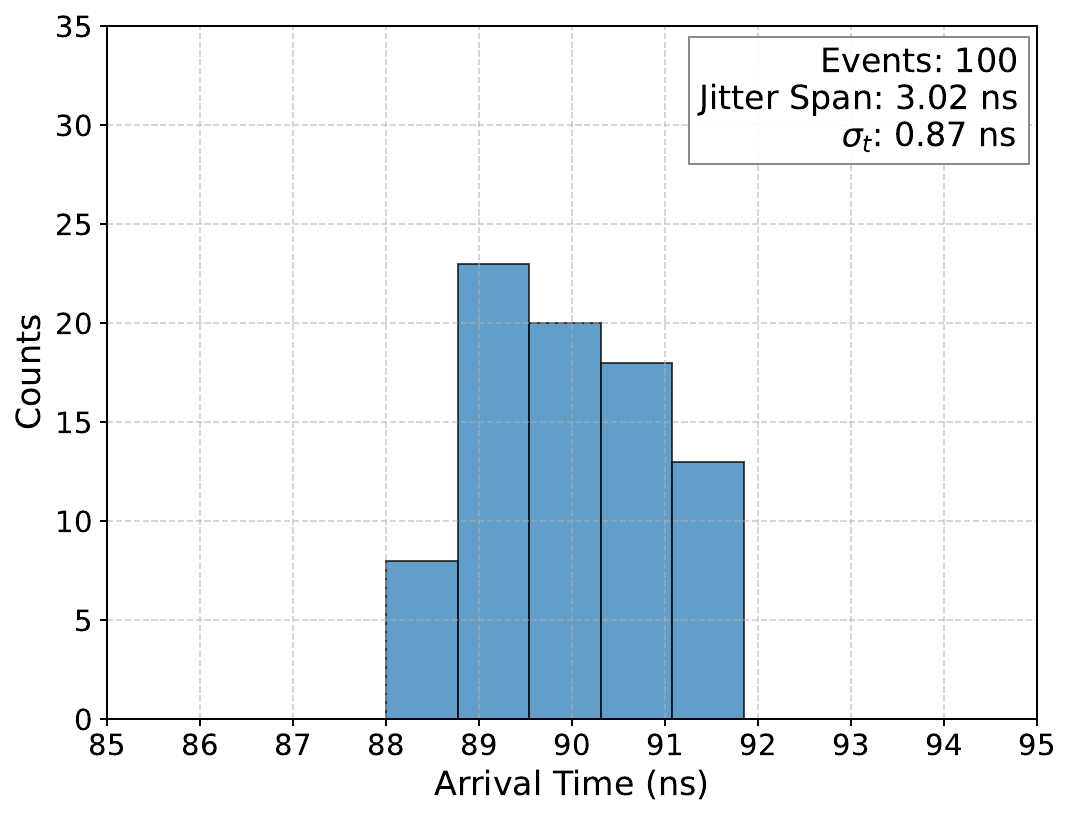}
    \caption{Distribution of the signal arrival times.}\label{fig:jitter_histogram}
\end{figure}

In the timing precision test, a signal generator was adopted to provide both the analog input and the trigger reference. The analog pulse was transmitted to the FEE, while the generator's marker signal was used as the system trigger. A fixed delay is maintained between the marker and the pulse output. Digitized waveforms from 100 consecutive trigger events were superimposed to analyze the temporal stability. The superposition of the waveform rising edges is shown in Fig.~\ref{fig:jitter_superposition}. As indicated by the discrete sample points (red dots), the captured rising edges are distributed within a specific time window. For the extraction of the precise signal arrival time ($t_{\text{arrival}}$), a linear interpolation method was applied at the 50\% threshold of the peak amplitude. Consequently, the quantization error inherent in discrete sampling is reduced, and the arrival times can be determined on a continuous scale. The observed timing spread is caused by the quantization uncertainty inherent in the synchronization process between the asynchronous external trigger and the internal $325\ \text{MHz}$ system clock. Consequently, the distribution of $t_{\text{arrival}}$ reflects the random phase difference between these two domains. The statistical distribution of the extracted arrival times is displayed in Fig.~\ref{fig:jitter_histogram}. The data shows a time span of $3.02\ \text{ns}$, which corresponds to approximately one cycle of the system clock. Due to the uniform nature of this distribution, the timing standard deviation $\sigma_t$ is derived using the formula $Span/\sqrt{12}$. This calculation yields a final timing precision of $\sigma_t = 0.87\ \text{ns}$.

In order to evaluate the contribution of this timing precision to the spatial resolution, a typical electron drift velocity of $v_d \approx 30$ $\mu$m/ns is assumed. Based on the measured $0.87\ \text{ns}$ timing jitter, the spatial uncertainty contribution is approximately $26.1\ \mu\text{m}$. The intrinsic spatial resolution of the drift chamber is typically around $130\ \mu\text{m}$. Compared to this value, the error introduced by the readout electronics is minor. The total resolution degradation is less than $2\%$ when combining these errors in quadrature. The result indicates that the intrinsic timing jitter of the readout system has a negligible impact on the position resolution capabilities of the detector.

\subsection{Statistical Characterization and Waveform Reconstruction}\label{subsec3.5}

\begin{figure}[H]
    \centering
    \includegraphics[width=0.98\linewidth]{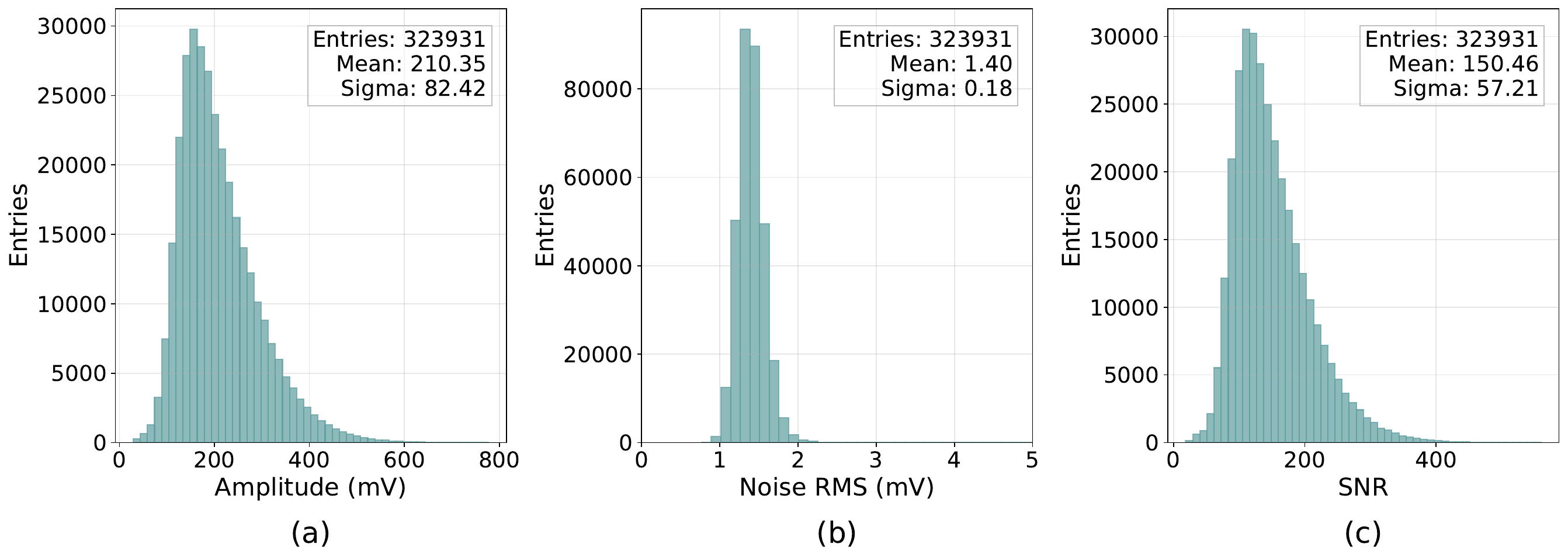}
    \caption{Statistical analysis of cosmic ray events: (a) Signal amplitude distribution; (b) RMS noise distribution; (c) Signal-to-Noise Ratio distribution.}\label{fig:snr_stats}
\end{figure}

To quantitatively assess the overall signal quality and noise performance, a statistical analysis was performed on a dataset of 323931 cosmic ray events. Valid single-channel waveforms were extracted from these events, with each waveform contributing a single entry to the statistical distributions. In this analysis, the amplitude is defined as the average height of all identified peaks within a single waveform, calculated after baseline subtraction. The distributions of signal amplitude, baseline noise, and the resulting Signal-to-Noise Ratio (SNR) are presented in Fig.~\ref{fig:snr_stats}. The signal amplitude distribution (Fig.~\ref{fig:snr_stats}(a)) exhibits a mean value of $210.35~\text{mV}$, while the noise RMS (Fig.~\ref{fig:snr_stats}(b)) remains low with a mean of $1.4~\mathrm{mV}_\text{rms}$. As shown in Fig.~\ref{fig:snr_stats}(c), the system achieves a mean SNR of $150.46$, which corresponds to a relative noise level of approximately $0.67\%$. This result confirms the low-noise characteristic and baseline stability of the readout system.

\begin{figure}[H]
  \centering
  \includegraphics[width=1\linewidth]{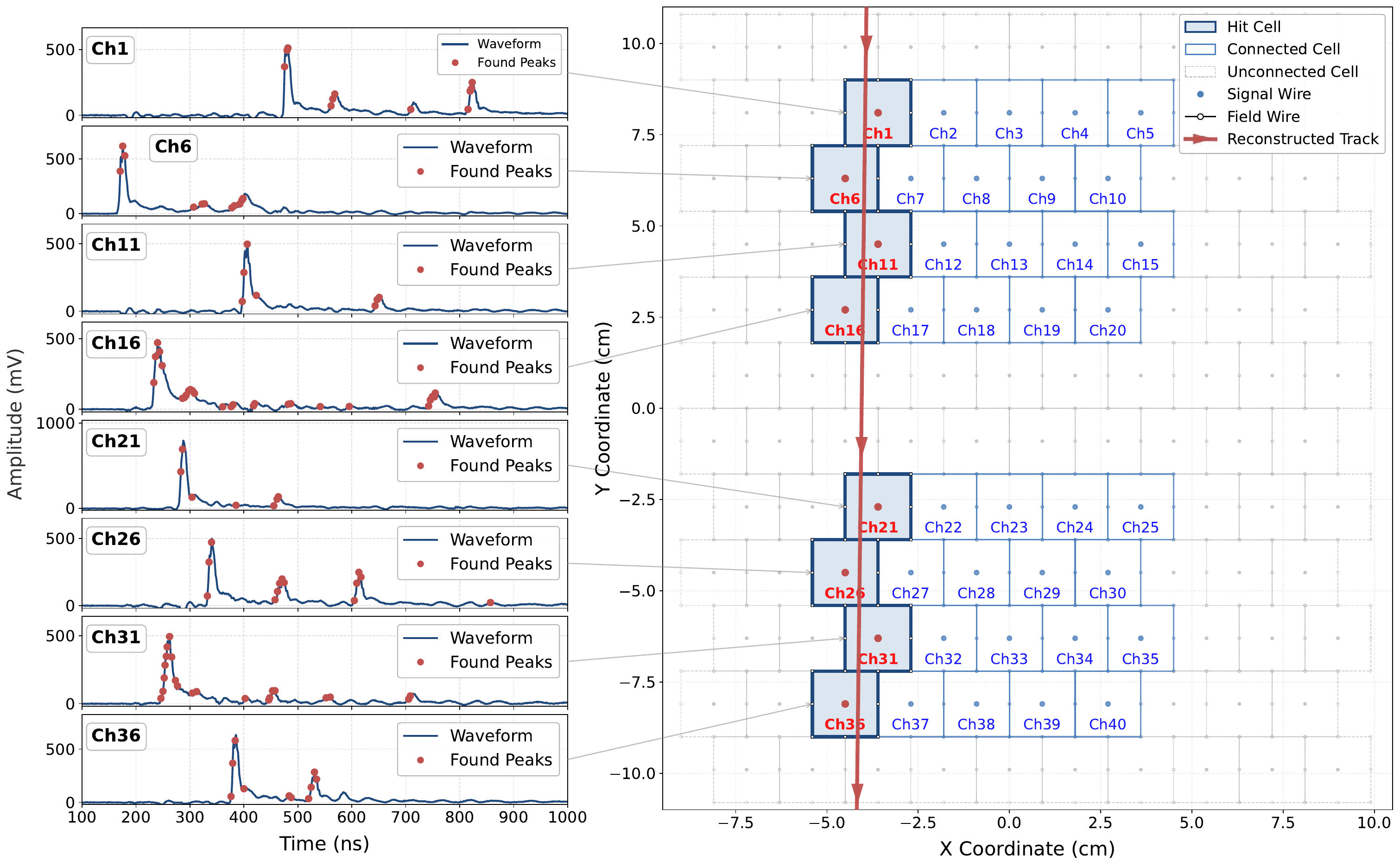}
  \caption{Typical cosmic-ray muon event captured by the drift chamber.}\label{fig:track_event}
\end{figure}

Following the statistical validation, the integrated system's functionality under realistic conditions was verified by acquiring cosmic-ray muon events in external trigger mode. A typical event captured by the drift chamber is shown in Fig.~\ref{fig:track_event}. The right plot depicts the geometric mapping of the hit cells, highlighted in blue, which display a linear spatial correlation consistent with the trajectory of a cosmic-ray muon. The digitized waveforms from the corresponding anode wires are shown in the left plot. Upon the arrival of the global trigger, valid signal peaks are observed across multiple channels within the readout time window. Although the arrival times vary due to different drift distances, the capture of these coincident signals confirms that the multi-channel synchronization logic functions correctly.

\begin{figure}[H]
  \centering
  \includegraphics[width=1\linewidth]{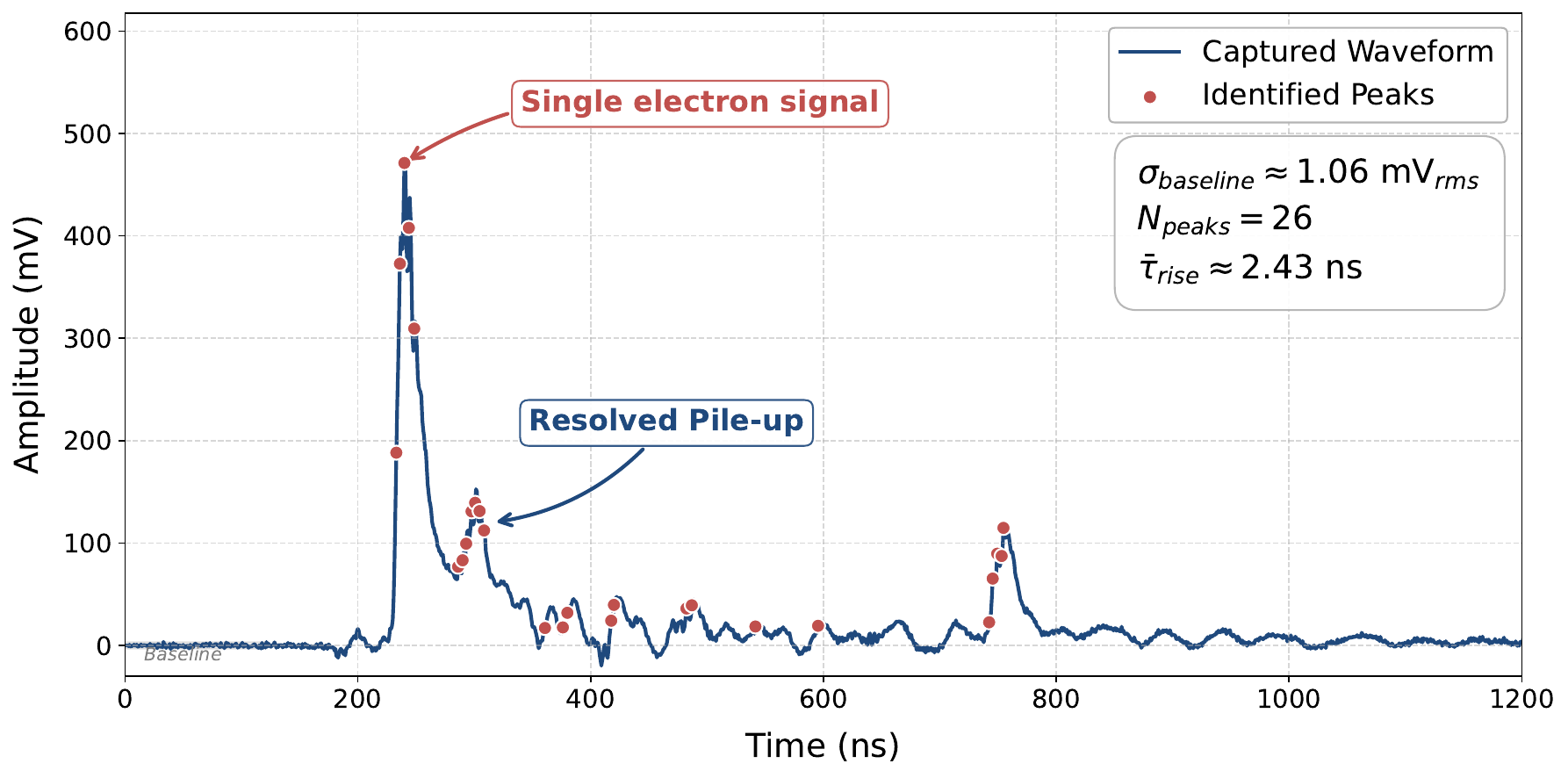}
  \caption{Waveform analysis of Channel 16 for primary ionization cluster resolution.}\label{fig:ch16_zoom}
\end{figure}

The resolution of discrete ionization peaks constitutes the basis for the $dN$ measurement. Consequently, the capability of the system to distinguish individual signal peaks was evaluated through a detailed waveform analysis of Channel 16 (shown in Fig.~\ref{fig:ch16_zoom}). The recorded signal demonstrates a stable baseline with an instantaneous noise of approximately $1.06~\mathrm{mV}_\text{rms}$ and an average rise time of $2.43~\text{ns}$. This fast transient response facilitates the discrimination of the temporal structure of electron clusters. As highlighted by the ``Resolved Pile-up'' annotation, the system distinguishes closely spaced peaks, indicating sufficient bandwidth to preserve the detailed ionization information. A preliminary second-derivative algorithm \cite{ZHAO2024109208} identifies the major peaks marked by red dots. While the quantitative correlation between these peaks and physical ionization clusters requires further calibration with beam tests, the high signal integrity observed confirms that the readout electronics provide a valid hardware platform capable of supporting cluster counting algorithms.

\section{Conclusion and Outlook }\label{sec4}
This paper presents the design, integration, and prototype characterization of a readout electronics system aimed at enabling \dNdX\ particle identification for the CEPC drift chamber. To meet the requirements of the cluster counting technique, the system incorporates a high-bandwidth analog front-end, a high-speed digitization module operating at $1.3\ \text{GSps}$,  and a waveform sampling-based timing unit that achieves a precision of $0.87\ \text{ns}$.

Through performance evaluations, the key hardware specifications of the system were verified. The measured results show that the $-3$~\text{dB} bandwidth of the readout chain is $460\ \text{MHz}$. The ENI is determined to be approximately $0.81~\mu\text{A}_{\text{rms}}$, satisfying the sensitivity requirement for detecting single primary electrons. Furthermore, statistical analysis of cosmic ray events demonstrates a relative noise level of $0.67\%$. These low-noise characteristics, combined with the high analog bandwidth, ensure the high sensitivity and fast dynamic response required to resolve closely spaced ionization signals. Regarding the timing performance, the intrinsic electronics jitter was measured to be $0.87~\text{ns}$. This corresponds to a negligible spatial contribution of $26.1~\mu\text{m}$ to the tracking resolution.

The preliminary cosmic-ray tests described in this paper have validated the architecture of the readout system and confirmed its ability to capture detailed waveform structures. These results demonstrate that the prototype hardware provides the necessary signal fidelity for ionization analysis. The ongoing work now focuses on completing the full integration of the 120-channel readout system. Once fully instrumented, further studies using MIPs will be conducted to perform a systematic characterization of the detector performance. Finally, beam tests are planned to quantitatively validate the cluster counting efficiency and evaluate the PID capability under realistic experimental conditions.

\bmhead{Acknowledgements}

This work was supported by the National Key Research and Development Program of China (Grant Nos. 2024YFE0110003 and 2024YFA1610604), the National Natural Science Foundation of China (Grant No. 12227810), Guangdong Basic and Applied Basic Research Foundation (Grant No. 2025A1515011426), and Guangdong Provincial Key Laboratory of Advanced Particle Detection Technology (Grant No. 2024B1212010005).

\section*{Declarations}

\bmhead{Conflict of interests} We declare that we have no financial and personal relationships with other people or organizations that can inappropriately influence our work, and there is no professional or other personal interest of any nature or kind in any product, service, and/or company that could be construed as influencing the position presented in, or the review of, the manuscript entitled “Integration and characterization of Readout Electronics System for dN/dx Measurement with Drift Chamber Prototype”.


\begin{thebibliography}{99}

\bibitem{thecepcstudygroup2025cepctechnicaldesignreport}
The CEPC Study Group, CEPC Technical Design Report -- Reference Detector (2025). arXiv:2510.05260 [hep-ex]

\bibitem{An_2019}
An, F., Bai, Y., Chen, C., et al.: Precision Higgs physics at the CEPC. Chin. Phys. C 43(4), 043002 (2019). doi:10.1088/1674-1137/43/4/043002

\bibitem{chekanov2016conceptual}
Chekanov, S.V., Demarteau, M.: Conceptual design studies for a CEPC detector. Int. J. Mod. Phys. A 31(33), 1644021 (2016)

\bibitem{Zhu2022RequirementAF}
Zhu, Y., Chen, S., Cui, H., Ruan, M.: Requirement analysis for dE/dx measurement and PID performance at the CEPC baseline detector. Nucl. Instrum. Methods Phys. Res. A 1047, 167835 (2023)

\bibitem{CHIARELLO2019464}
Chiarello, G., Chiri, C., Cocciolo, G., et al.: Improving spatial and PID performance of the high transparency Drift Chamber by using the Cluster Counting and Timing techniques. Nucl. Instrum. Methods Phys. Res. A 936, 464--465 (2019)

\bibitem{article}
Liu, M.Y., Li, W.D., Huang, X.T., et al.: Simulation and reconstruction of particle trajectories in the CEPC drift chamber. Nucl. Sci. Tech. 35, 128 (2024). doi:10.1007/s41365-024-01497-z

\bibitem{Huang2025}
Huang, F., Dong, M., Liu, H., et al.: Experimental study of primary ionization counting method. Radiat. Detect. Technol. Methods (2025). doi:10.1007/s41605-025-00572-2

\bibitem{Fang_2023}
Fang, W., Huang, X., Li, W., et al.: Precise simulation of drift chamber in the CEPC experiment. J. Phys.: Conf. Ser. 2438, 012113 (2023). doi:10.1088/1742-6596/2438/1/012113

\bibitem{Chiarello_2017}
Chiarello, G., Chiri, C., Cocciolo, G., et al.: Application of the Cluster Counting/Timing techniques to improve the performances of high transparency Drift Chamber for modern HEP experiments. J. Instrum. 12(07), C07021 (2017). doi:10.1088/1748-0221/12/07/C07021

\bibitem{dc_electronics}
Panareo, M., et al.: The front end electronics for the drift chamber readout in MEG experiment upgrade. J. Instrum. 15(07), C07009 (2020)

\bibitem{cluster_ml}
Tian, Z., et al.: Cluster Counting Algorithm for Drift Chamber using LSTM and DGCNN. arXiv preprint arXiv:2402.16493 (2024)

\bibitem{lmh6629}
Texas Instruments, Ultra-low-noise high-speed operational amplifier with shutdown. \url{https://www.ti.com/product/LMH6629}

\bibitem{ad9695}
Analog Devices, 14-Bit, 1300 MSPS/625 MSPS, JESD204B, Dual Analog-to-Digital Converter. \url{https://www.analog.com/en/products/ad9695.html}

\bibitem{lmk04828}
Texas Instruments, Ultra low-noise JESD204B compliant clock jitter cleaner. \url{https://www.ti.com/product/LMK04828}

\bibitem{zynq-ultrascale}
Advanced Micro Devices, AMD Zynq UltraScale+ MPSoCs. \url{https://www.amd.com/en/products/adaptive-socs-and-fpgas/soc/zynq-ultrascale-plus-mpsoc.html}

\bibitem{ZHAO2024109208}
Zhao, G., Wu, L., Grancagnolo, F., et al.: Peak finding algorithm for cluster counting with domain adaptation. Comput. Phys. Commun. 300, 109208 (2024). doi:10.1016/j.cpc.2024.109208

\end{thebibliography}
\end{document}